\documentclass[conference]{IEEEtran}
\IEEEoverridecommandlockouts
\usepackage{cite}
\usepackage{amsmath,amssymb,amsfonts}
\usepackage[acronym]{glossaries}
\usepackage{algorithmic}
\usepackage{authblk}
\usepackage[pdftex]{graphicx}
\usepackage{subcaption}
\usepackage[T1]{fontenc}
\usepackage{newtxtext,newtxmath}
\usepackage{epstopdf}
\usepackage{textcomp}
\usepackage{xcolor}
\def\BibTeX{{\rm B\kern-.05em{\sc i\kern-.025em b}\kern-.08em
    T\kern-.1667em\lower.7ex\hbox{E}\kern-.125emX}}
\begin{document}

\title{Advanced Channel Decomposition Techniques in OTFS: A GSVD Approach for Multi-User Downlink\\
% {\footnotesize \textsuperscript{*}Note: Sub-titles are not captured in Xplore and
% should not be used}
\thanks{This work was conducted within the MiFuture project, which has received funding from the European Union's Horizon Europe (HE) Marie Skłodowska-Curie Actions Doctoral Networks (MiFuture HORIZON-MSCA-2022-DN-01 and YAHYA/6G HORIZON-MSCA-2022-PF-01) under Grant Agreement number 101119643. It was partially supported by Fundação para a Ciência e Tecnologia and Instituto de Telecomunicações under the project UIDB/50008/2020 DOI:10.54499/UIDB/50008/2020.}
}

% \author{\IEEEauthorblockN{1\textsuperscript{st} Given Name Surname}
% \IEEEauthorblockA{\textit{dept. name of organization (of Aff.)} \\
% \textit{name of organization (of Aff.)}\\
% City, Country \\
% email address or ORCID}
% \and
% \IEEEauthorblockN{2\textsuperscript{nd} Given Name Surname}
% \IEEEauthorblockA{\textit{dept. name of organization (of Aff.)} \\
% \textit{name of organization (of Aff.)}\\
% City, Country \\
% email address or ORCID}
% \and
% \IEEEauthorblockN{3\textsuperscript{rd} Given Name Surname}
% \IEEEauthorblockA{\textit{dept. name of organization (of Aff.)} \\
% \textit{name of organization (of Aff.)}\\
% City, Country \\
% email address or ORCID}
% \and
% \IEEEauthorblockN{4\textsuperscript{th} Given Name Surname}
% \IEEEauthorblockA{\textit{dept. name of organization (of Aff.)} \\
% \textit{name of organization (of Aff.)}\\
% City, Country \\
% email address or ORCID}
% \and
% \IEEEauthorblockN{5\textsuperscript{th} Given Name Surname}
% \IEEEauthorblockA{\textit{dept. name of organization (of Aff.)} \\
% \textit{name of organization (of Aff.)}\\
% City, Country \\
% email address or ORCID}
% \and
% \IEEEauthorblockN{6\textsuperscript{th} Given Name Surname}
% \IEEEauthorblockA{\textit{dept. name of organization (of Aff.)} \\
% \textit{name of organization (of Aff.)}\\
% City, Country \\
% email address or ORCID}
% }
\author[1,2]{Omid Abbassi Aghda}
\author[1,3]{Oussama Ben Haj Belkacem} 
\author[4]{Dou Hu} 
\author[1,2]{João Guerreiro} 
\author[1,5]{Nuno Souto} 
\author[2,6]{\\Michal Szczachor} 
\author[1,2]{Rui Dinis} \affil[1]{Instituto de Telecomunicações, Lisboa, Portugal} \affil[2]{Universidade Nova de Lisboa, Monte da Caparica, 2829-516 Caparica, Portugal} 
\affil[3]{Innov’Com Laboratory, Sup’Com, University of Carthage, Tunis 1054, Tunisia} 
\affil[4]{University of Tokyo, Tokyo, Japan}
\affil[5]{ ISCTE-Instituto Universitário de Lisboa,
1649-026 Lisbon, Portugal}
\affil[6]{Nokia, Wroclaw, Poland}
\newacronym{qam}{QAM}{quadrature amplitude modulation}
\newacronym{dd}{DD}{delay-Doppler}
\newacronym{otfs}{OTFS}{orthogonal time frequency space}
\newacronym{noma}{NOMA}{non-orthogonal multiple access}
\newacronym{mmse}{MMSE}{minimum mean square error}
\newacronym{snr}{SNR}{signal to noise ratio}
\newacronym{bd}{BD}{block diagonalization}
\newacronym{ber}{BER}{bit error rate}
\newacronym{2d}{2D}{ 2 dimensional}
\newacronym{ofdm}{OFDM}{Orthogonal Frequency Division Multiplexing}
\newacronym{dl}{DL}{deep learning}
\newacronym{bs}{BS}{base station}
\newacronym{gsvd}{GSVD}{generalized singular value decomposition}
\newacronym{hogsvd}{HO-GSVD}{Higher Order GSVD}
\newacronym{ia}{IA}{interference alignment}
\newacronym{iui}{IUI}{inter user interference}
\newacronym{mu}{MU}{multi user}
\newacronym{svd}{SVD}{singular value decomposition}
\newacronym{mimo}{MIMO}{multiple input multiple output}
\newacronym{siso}{SISO}{single input single output}
\newacronym{cc}{CC}{common channel}
\newacronym{pc}{PC}{private channel}
\newacronym{oma}{OMA}{Orthogonal multiple access}
\newacronym{sic}{SIC}{sucessive interference cancellation}
\newacronym{tf}{TF}{time-frequency}
\newacronym{mmse}{MMSE}{minimum mean square error}
\newacronym{zf}{ZF}{zero forcing}
\newacronym{ma}{MA}{multiple access}
\newacronym{iai}{IAI}{inter-antenna interference}
\newacronym{fde}{FDE}{frequency domain equalization}
\newacronym{scm}{SCM}{single carrier modulation}
\newacronym{jm}{JM}{joint matrix}
\newcommand{\RNum}[1]{\uppercase\expandafter{\romannumeral #1\relax}}
\maketitle

\begin{abstract}
In this paper, we propose a multi-user downlink system for two users based on the orthogonal time frequency space (OTFS) modulation scheme. The design leverages the generalized singular value decomposition (GSVD) of the channels between the base station and the two users, applying precoding and detection matrices based on the right and left singular vectors, respectively. We derive the analytical expressions for three scenarios and present the corresponding simulation results. These results demonstrate that, in terms of bit error rate (BER), the proposed system outperforms the conventional multi-user OTFS system in two scenarios when using minimum mean square error (MMSE) equalizers or precoder, both for perfect channel state information and for a scenario with channel estimation errors. In the third scenario, the design is equivalent to zero-forcing (ZF) precoding at the transmitter.
\end{abstract}


\section{Introduction}
Reliable communication in the sixth generation (6G) of wireless networks is crucial for high-speed users. \Gls{otfs} modulation is specifically designed for this scenario, as it multiplexes information in the \gls{dd} domain, enabling it to track both the delay and the Doppler spread of the channel \cite{Hadani7925924,Raviteja8424569}. However, several challenges must be addressed when dealing with \gls{ma} in \Gls{otfs}.

\gls{oma} methods in \gls{otfs} modulation have been explored in various studies. The authors in \cite{Rakib2017} proposed multiplexing in the \gls{dd} domain and allocating different \gls{dd} bins to different users. This method is simple on the transmitter side; however, each user must cancel out interference from other users at the receiver. To address this, guard intervals can be considered between the multiplexed data in the \gls{dd} domain, which results in lower spectral efficiency. In \cite{Khammammetti8515088}, \gls{tf} resource allocation is considered for \gls{otfs} modulation, where information symbols are multiplexed and interleaved in the \gls{dd} domain so that their corresponding \gls{tf} domain is contiguous.
In the study \cite{Reddy10318840}, an uplink \gls{ma} scenario was proposed based on \gls{ofdm} modulation for \gls{siso} systems. They demonstrated that the concatenated signal from all symbols at the \gls{bs} is equivalent to a single-user \gls{otfs} system. Their work was further extended in a subsequent study \cite{Sudhakar10725940}, where they investigated the uplink scenario for \gls{mimo}

Similarly, \gls{noma} has been explored in the context of \gls{otfs} modulation. The authors in \cite{Umakoglu10646342} proposed a deep learning-based signal detection method for a two-user downlink \gls{siso}-\gls{noma} system. In \cite{Ding8901184},  a robust beamforming method is proposed for a downlink \gls{mimo} system with two users. This method multiplexes the low mobility user's information in the \gls{tf} domain while ensuring the high mobility user meets the minimum requirements by multiplexing its information in the \gls{dd} domain. In \cite{Yang10634195}, the authors utilize an interference alignment (IA) matrix as a precoder to reduce inter-user interference (IUI) in a \gls{mu} \gls{mimo}-\gls{otfs} downlink system. They also consider singular value decomposition (SVD) and precoding to diagonalize the channel matrix and perform data detection.

The other remaining challenge in \gls{otfs} modulation is the equalization, despite much research on the subject. The input-output equation of the transceiver in the \gls{dd} domain is modeled as a \gls{2d} phase-rotated convolution between the transmitted signal and the channel impulse response in the \gls{dd} domain \cite{Raviteja8424569}. Due to this \gls{2d} convolution, and compared to \gls{ofdm}, which allows for a low-complexity single-tap equalizer, \gls{otfs} channel equalization and data detection remains challenging due to high computational complexity. In high-speed scenarios, \gls{fde} is ineffective for \gls{otfs}. Unlike in low-mobility scenarios with \gls{scm}, where a single-tap equalizer can be used for \gls{fde}, this approach does not work for \gls{otfs}. It is worth mentioning that the performance of the \gls{otfs} system is not affected by user speed, unlike high mobility system designs in \gls{ofdm} and \gls{scm}, if an adequate equalization or detection technique is applied.

The \gls{gsvd} decomposition was initially proposed in \cite{van1976generalizing}, with a more comprehensive definition provided in \cite{paige1981towards}. Additionally, the authors in \cite{hogsvd} introduced the \gls{hogsvd} to decompose more than two matrices. In \cite{Senaratne6415345}, the author utilized \gls{gsvd} to separate a wireless channel for two users into the \gls{pc} and \gls{cc}, offering a complete overview of both definitions of the \gls{gsvd} decomposition.

Our contribution in this paper is the development of a precoding and detection scheme for a \gls{mu} \gls{mimo}-\gls{otfs} system based on \gls{gsvd} decomposition. This decomposition allows us to reduce the channel matrix to a diagonal form, thereby avoiding the need to handle a \gls{2d} circular convolution at the receiver. We then evaluate the performance of the proposed system and compare it with other \gls{mu} \gls{mimo} systems, including \gls{bd} precoding, \gls{mmse} precoding, and \gls{mmse} equalization. Our proposed method demonstrates superiority over these mentioned methods.

\textbf{Notation}: In the rest of this paper Boldface upper case, Boldface lower case, and normal lower case indicate matrices, vectors, and scalars, respectively. Notations $(.)^{H}$, $(.)^T$, and $\text{vec}{.}$ are used to represent the Hermitian, transpose, and column-wise vectorization. The notation $\mathbf{A}_{\left\{ :,1:k \right\}}$ indicates the selection of $1^{st}$ to the $k^{th}$ column of the matrix $\mathbf{A}$. 
$\mathbf{I}_k$, $\mathbf{0}_k$, and $\mathbf{0}_{k \times r}$ denote the identity matrix of size $k \times k$,  the zero matrix of size $k \times k$, and the zero matrix of size $k \times r$, respectively.

\section{OTFS  system model}
In \gls{otfs} modulation, the \gls{dd} domain is discretized into an $M \times N$ grid, where $M$ and $N$ are the number of bins along the delay and Doppler axes, respectively. In this grid, the $l^{th}$ bin ($l = 0, 1, \ldots, M-1$) corresponds to a delay of $\tau = \frac{l}{B}$, and the $k^{th}$ bin ($k = 0, 1, \ldots, N-1$) corresponds to a Doppler shift of $\nu = \frac{k}{T_f}$. Here, $B$ and $T_f$ represent the bandwidth and the frame duration, respectively. Accordingly, the sub-carrier spacing for the corresponding \gls{tf} domain and the \gls{otfs} sub-symbol duration are $\Delta f = \frac{B}{M}$ and $T_s = \frac{T_f}{N}$, respectively, such that $T_s \Delta f = 1$.

\subsection{SISO System model}
In \gls{siso}-\gls{otfs}, information precoded symbols are multiplexed in the \gls{dd} grid, represented by the matrix $\mathbf{X} \in \mathbb{C}^{M \times N}$. This signal passes through the equivalent \gls{dd} channel representation, and the received signal in the \gls{dd} domain is given by \cite[Chapter 4]{hong2022delay}
\begin{equation} 
\mathbf{y} = \mathbf{Hx} + \mathbf{n}, 
\end{equation} 
where $\mathbf{x} = \text{vec}\left\{\mathbf{X}^T\right\}\in\mathbb{C}^{MN\times 1}$ and $\mathbf{y} = \text{vec}\left\{\mathbf{Y}^T\right\}\in\mathbb{C}^{MN\times 1}$, with $\mathbf{Y}\in\mathbb{C}^{M\times N}$ being the \gls{dd} representation of the received signal. The channel matrix $\mathbf{H} \in \mathbb{C}^{MN \times MN}$ represents the equivalent \gls{dd} channel between $\mathbf{x}$ and $\mathbf{y}$.
 
\subsection{MIMO MA System Model}
The same \gls{dd} domain discretization used for \gls{siso} is applied to the \gls{mimo} case. We consider a downlink system with $C$ antennas at the base station and 2 users, each equipped with $G$ antennas. For simplicity, we focus on 2 users in this paper.\footnote{While this work can be  extended to multiple users with different numbers of antennas using \gls{hogsvd} \cite{hogsvd} \cite{Senaratne6415345}.
}
 The signal representation in the \gls{dd} domain for the $c^{th}$ antenna is $\mathbf{X}_c \in \mathbb{C}^{M \times N}$ for $c = 1, 2, \ldots, C$.

The received signal at the $g^{th}$ antenna for $1^{st}$ and $2^{nd}$ user is given by
\begin{equation}
    \begin{cases}
        \mathbf{y}_{g,1} = \mathbf{H}_{g,1,1}\mathbf{x}_1 + \mathbf{H}_{g,1,2}\mathbf{x}_2+ \cdots +\mathbf{H}_{g,1, C}\mathbf{x}_{C} +\mathbf{n}_{g,1}  \\
        \mathbf{y}_{g,2} = \mathbf{H}_{g,2,1}\mathbf{x}_1 + \mathbf{H}_{g,2,2}\mathbf{x}_2+ \cdots +\mathbf{H}_{g,2, C}\mathbf{x}_{C} +\mathbf{n}_{g,2},  
    \end{cases}   
    \label{eq:input-output for each antenna of each user}
\end{equation}
where $\mathbf{y}_{g,1}, \mathbf{y}_{g,2} \in \mathbb{C}^{MN \times 1}$ for $g = 1, 2, \ldots, G$. In \eqref{eq:input-output for each antenna of each user}, $\mathbf{H}_{g,1,c} \in \mathbb{C}^{MN \times MN}$ and $\mathbf{H}_{g,2,c} \in \mathbb{C}^{MN \times MN}$ are the \gls{dd} channel matrices between the $c^{th}$ antenna of the transmitter and the $g^{th}$ antenna of the receiver for the $1^{st}$ and $2^{nd}$ user, respectively. For this paper, we assume a rich physical environment characterized by a high number of multipath components. This ensurs that the matrices $\mathbf{H}_{g,1,c}$ and $\mathbf{H}_{g,2,c}$ are full-rank, i.e., $\text{rank}(\mathbf{H}_{g,1,c}) = \text{rank}(\mathbf{H}_{g,2,c}) = MN$.
The vector $\mathbf{x}_c$ for $c = 1, 2, \ldots, C$ is the transmitted signal vector, such that $\mathbf{x}_c = \text{vec}(\mathbf{X}_c^T) \in \mathbb{C}^{MN\times 1}$.

The received signals from all antennas for each user can be concatenated, allowing \eqref{eq:input-output for each antenna of each user} to be rewritten as
\begin{equation}
    \begin{cases}
        \mathbf{y}_1 = \mathbf{H}_1 \mathbf{x}+ \mathbf{n}_1  \\
        \mathbf{y}_2 = \mathbf{H}_2 \mathbf{x}+ \mathbf{n}_2. 
    \end{cases}  
    \label{eq:input output for each user}
\end{equation}
The received vector for the $1^{st}$ user, $\mathbf{y}_1$, and the transmitted vector from the \gls{bs}, $\mathbf{x}$, are given by
\begin{equation}
     \mathbf{y}_1 = \begin{bmatrix}
        \mathbf{y}_{1,1}\\
        \mathbf{y}_{2,1}\\
        \vdots\\
        \mathbf{y}_{G,1}
\end{bmatrix}\in \mathbb{C}^{MNG\times 1}, 
\mathbf{x} = \begin{bmatrix}
        \mathbf{x}_{1}\\
        \mathbf{x}_{2}\\
        \vdots\\
        \mathbf{x}_{C}
\end{bmatrix}\in \mathbb{C}^{MNC\times 1}.
\end{equation}
The channel matrix between the \gls{bs} and the $1^{th}$ user is modeled as
\begin{equation}
\begin{split}
\mathbf{H}_1=
\begin{bmatrix}
    \mathbf{H}_{1,1,1} & \mathbf{H}_{1,1,2} & \cdots & \mathbf{H}_{1,1,C} \\
    \mathbf{H}_{2,1,1} & \mathbf{H}_{2,1,2} & \cdots & \mathbf{H}_{2,1,C} \\
    \vdots & \ddots & \ddots & \vdots \\
    \mathbf{H}_{G,1,1} & \mathbf{H}_{G,1,2} & \cdots & \mathbf{H}_{G,1,C}    
\end{bmatrix} \\
\in \mathbb{C}^{MNG\times MNC}.
\end{split}
\end{equation}
Similarly, $\mathbf{y}_2$ and $\mathbf{H}_2$ represent the received vector and channel matrix for the $2^{nd}$ user.
 \section{GSVD decomposition, Precoding and Detection Matrix}
 \begin{figure*}[!t]
    \centering
    \begin{subfigure}[b]{0.25\textwidth}
        \includegraphics[width=\textwidth]{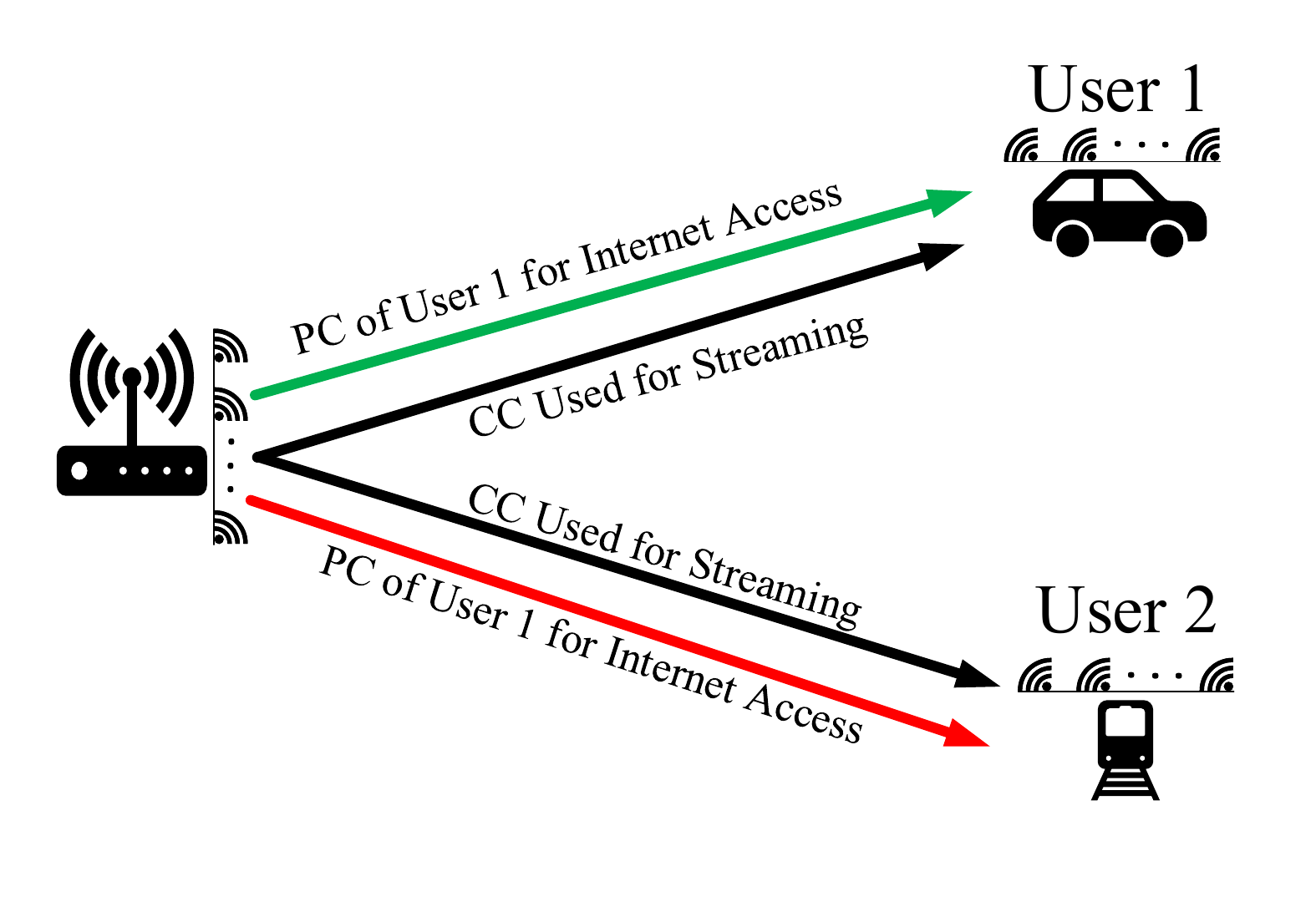}
        \caption{First Scenario, $G\leq C\leq2G$}
        \label{fig:sub1}
    \end{subfigure}
    \hfill
    \begin{subfigure}[b]{0.25\textwidth}
        \includegraphics[width=\textwidth]{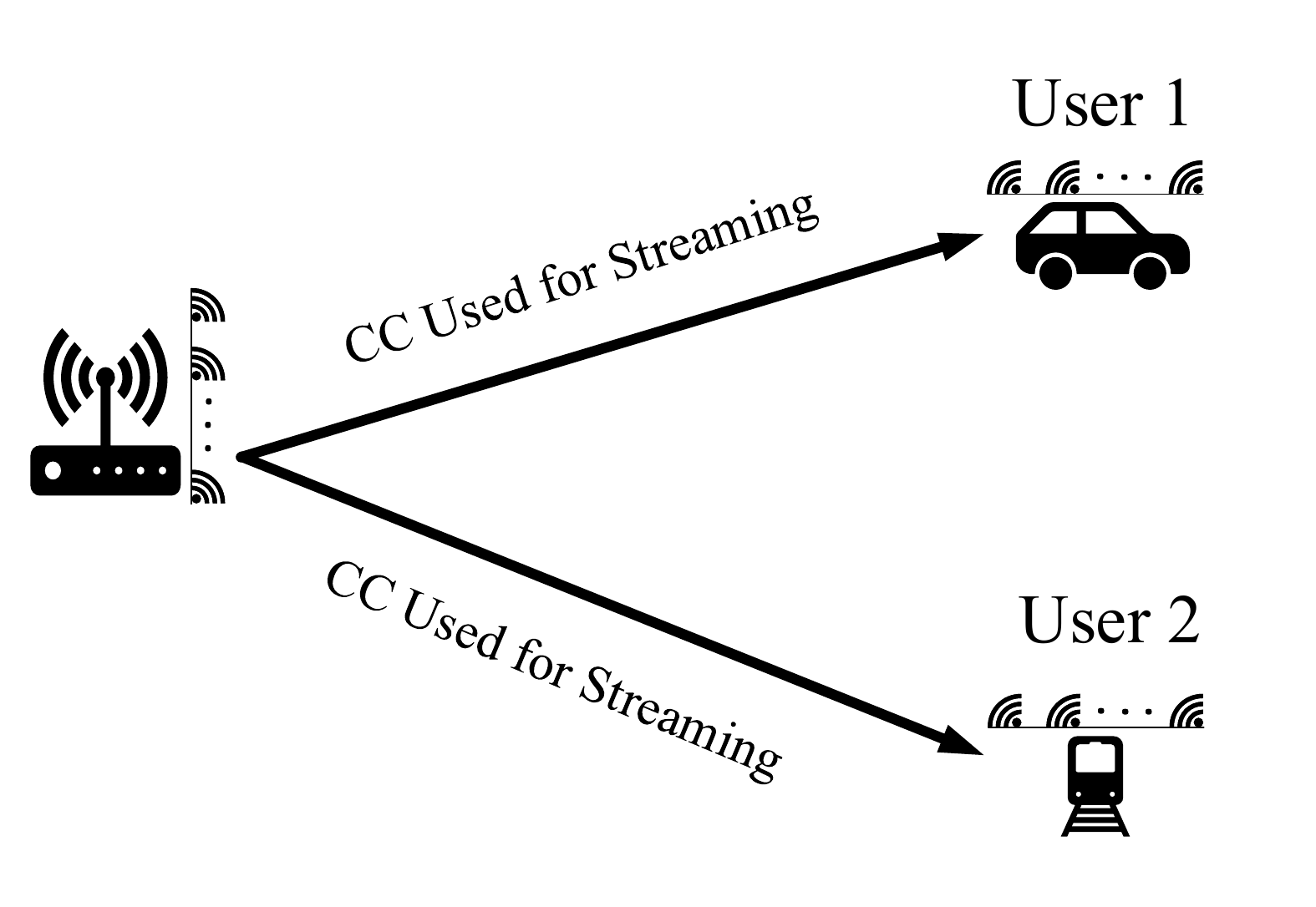}
        \caption{Second  Scenario, $C\leq G$}
        \label{fig:sub2}
    \end{subfigure}
    \hfill
    \begin{subfigure}[b]{0.25\textwidth}
        \includegraphics[width=\textwidth]{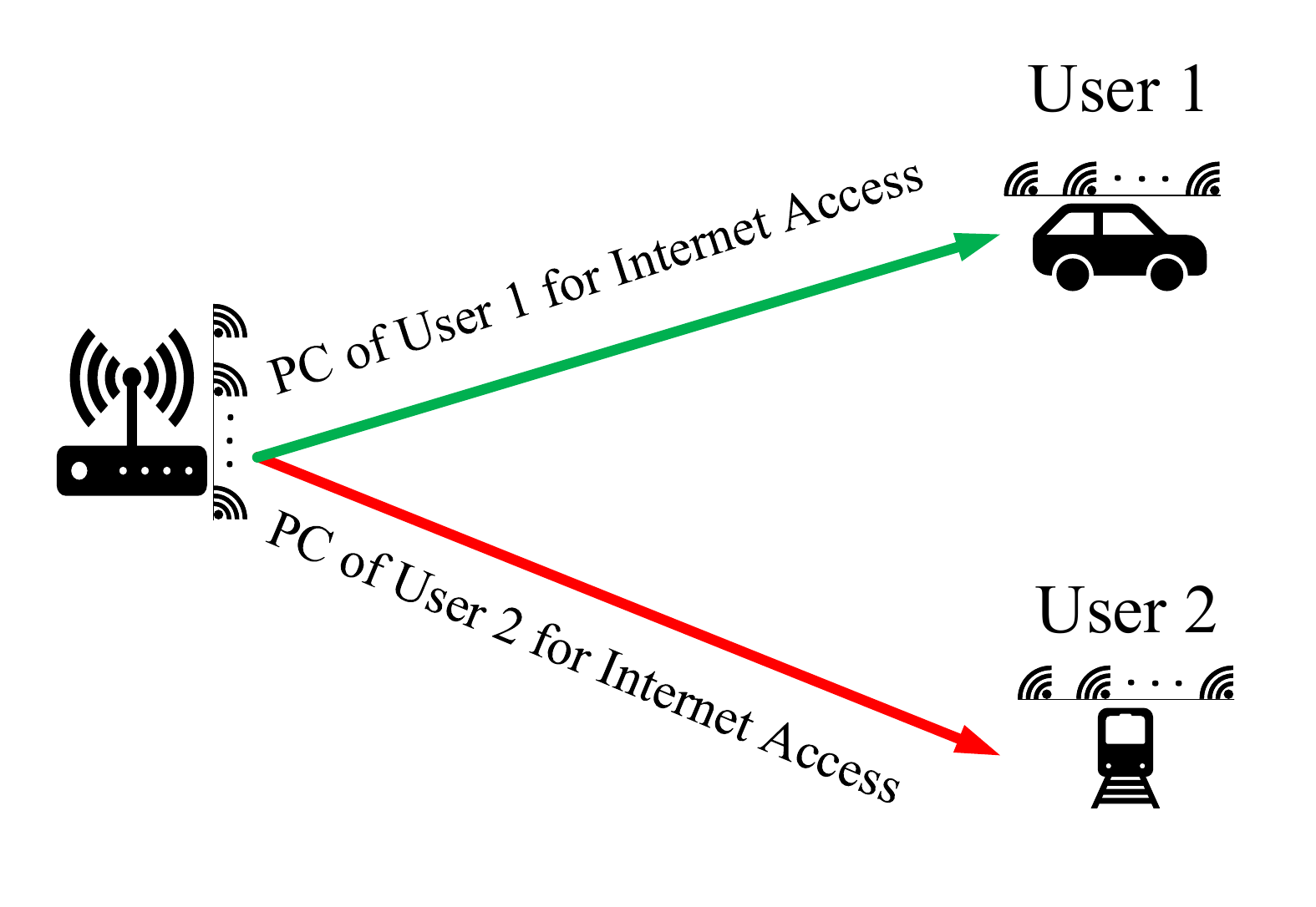}
        \caption{Third Scenario, $ 2G\leq C$}
        \label{fig:sub3}
    \end{subfigure}
    \caption{Schematic of different scenarios in GSVD-based channel decomposition for precoding and detection matrix design.
}
    \label{fig:overall}
\end{figure*}

The precoding matrix $\mathbf{P} \in \mathbb{C}^{MNC \times MNs}$ is defined such that $\mathbf{x} = \frac{1}{\rho}\mathbf{P}\mathbf{s}$. In this context, $s = \text{rank}\{\mathbf{H}\}/(MN)$, where $\mathbf{H} = [\mathbf{H}_1^T, \mathbf{H}_2^T]^T$, representing the number of data streams. The scalar $\rho$ is a normalization factor used to ensure constant average power at the transmitter\footnote{The calculation of the normalization factor $\rho$ is fully explained in \cite{Senaratne6415345}.}.  Vector $\mathbf{s} \in \mathbb{C}^{MNs \times 1}$ denotes the information \gls{qam} symbols.
In the receiver, we use the vectors $\mathbf{r}_1 = \rho\mathbf{D}_1\mathbf{y}_1$ and $\mathbf{r}_2 = \rho\mathbf{D}_2\mathbf{y}_2$, with $\mathbf{r}_1, \mathbf{r}_2 \in \mathbb{C}^{MNG \times 1}$, as the signal for detection purposes. The detection matrices for the first and second users are $\mathbf{D}_1, \mathbf{D}_2 \in \mathbb{C}^{MNG \times MNG}$, respectively. We define the precoding and detection matrices in the following  of this section.

Based on the \gls{gsvd}, the channel matrices $\mathbf{H}_1$ and $\mathbf{H}_2$ are decomposed as
 \begin{equation}
     \begin{cases}
         \mathbf{H}_1 = \mathbf{U}_1\mathbf{\Sigma}_1\mathbf{V}^H\\     
         \mathbf{H}_2 = \mathbf{U}_2\mathbf{\Sigma}_2\mathbf{V}^H.
     \end{cases}
   \label{eq:gsvd decomposition}  
 \end{equation}

Here, $\mathbf{U}_1, \mathbf{U}_2 \in \mathbb{C}^{MNG \times MNG}$ are unitary matrices, $\mathbf{\Sigma}_1, \mathbf{\Sigma}_2 \in \mathbb{C}^{MNG \times MNs}$ are block diagonal matrices, and $\mathbf{V} \in \mathbb{C}^{MNC \times MNs}$ is the \gls{jm}.

% We consider three scenarios:
% \begin{enumerate}
%     \item The number of transmitter antennas is greater than the number of antennas for each user but less than the total number of antennas for all users, i.e., $G \leq C \leq 2G$.
%     \item The total number of receiver antennas is less than the number of transmitter antennas, i.e., $2G \leq C$.
%     \item The number of transmitter antennas is smaller than the number of antennas for each user, i.e., $C \leq G$.
% \end{enumerate}

We consider three scenarios: \( G \leq C \leq 2G \), \( C \leq G \), and \( 2G \leq C \).
The detection matrices for each scenario are $\mathbf{D}_1 = \mathbf{U}_1^H$ and $\mathbf{D}_2 = \mathbf{U}_2^H$. Moreover, the structures of $\mathbf{\Sigma}_1$, $\mathbf{\Sigma}_2$, and $\mathbf{V}$, and consequently the precoding matrix $\mathbf{P}$, are explained in each scenario.
\subsection{Scenario \RNum{1}: $G\leq C\leq2G$}

In this case, we have $s = C$ data streams, and $\mathbf{V}$ is a square invertible matrix. Therefore, we choose the precoding matrix as $\mathbf{P} = (\mathbf{V}^H)^{-1}$. By substituting $\mathbf{P}$, $\mathbf{D}_1$, $\mathbf{D}_2$, and the \gls{gsvd} decomposition of \eqref{eq:gsvd decomposition} into \eqref{eq:input output for each user}, we obtain
\begin{equation}
    \begin{cases}
        \mathbf{r}_1=\mathbf{U}_1^H\mathbf{y}_1 = \mathbf{U}_1^H\mathbf{U}_1\mathbf{\Sigma}_1\mathbf{V}^H\left(\mathbf{V}^H\right)^{-1}\mathbf{s}+ \mathbf{U}_1^H\mathbf{n}_1\\
        \mathbf{r}_1=\mathbf{U}_2^H\mathbf{y}_2 = \mathbf{U}_2^H\mathbf{U}_2\mathbf{\Sigma}_1\mathbf{V}^H\left(\mathbf{V}^H\right)^{-1}\mathbf{s} + \mathbf{U}_2^H\mathbf{n}_2.
    \end{cases}
    \label{eq:gsvd replacement in the input output equation}
\end{equation}

Here,

\begin{equation}
    \mathbf{\Sigma}_1 =
    \begin{bmatrix}
        \mathbf{I}_{r} & \mathbf{0}_{r\times t} & \mathbf{0}_{r} \\
        \mathbf{0}_{t\times r} & \mathbf{C}_1 & \mathbf{0}_{t\times r}
    \end{bmatrix}, 
    \mathbf{\Sigma}_2  =
    \begin{bmatrix}
        \mathbf{0}_{t\times r} & \mathbf{C}_2 & \mathbf{0}_{t\times r} \\
        \mathbf{0}_{r} & \mathbf{0}_{r\times t} & \mathbf{I}_r
    \end{bmatrix},
    \label{eq:scenario_1_sigma}
\end{equation}
where $r = MN(C - G)$, $t = MN(2G - C)$, and $\mathbf{C}_1, \mathbf{C}_2 \in \mathbb{C}^{t \times t}$ are diagonal matrices with the \gls{cc} coupling coefficients for user 1 and user 2, respectively. Matrix $\mathbf{I}_{r}$ represents the \gls{pc} for the first and second users. Since the detection matrices $\mathbf{D}_1$ and $\mathbf{D}_2$ are unitary, there is no noise enhancement in the receiver due to the terms $\mathbf{U}_1^H \mathbf{n}_1$ and $\mathbf{U}_2^H \mathbf{n}_2$.
Moreover, equation \eqref{eq:gsvd replacement in the input output equation} can be simplified to
\begin{equation}
    \begin{cases}
        \mathbf{r}_1 =\mathbf{\Sigma}_1\mathbf{s}+ \mathbf{n}_1^{'}\\
        \mathbf{r}_2 = \mathbf{\Sigma}_2\mathbf{s} + \mathbf{n}_2^{'},
    \end{cases}
    \label{eq:gsvd diagonal channel input output equation}
\end{equation}
where $\mathbf{n}_1^{'} = \mathbf{U}_1^H \mathbf{n}_1$ and $\mathbf{n}_2^{'} = \mathbf{U}_2^H \mathbf{n}_2$.
Since in \eqref{eq:gsvd diagonal channel input output equation} the matrices $\mathbf{\Sigma}_1$ and $\mathbf{\Sigma}_2$ are diagonal, the information symbols $\mathbf{s}$ can be easily detected with a single tap \gls{mmse} equalizer.

Observing carefully \eqref{eq:gsvd diagonal channel input output equation} and comparing it with \eqref{eq:scenario_1_sigma}, it can be seen that users 1 and 2 receive $r$ private data streams through their own \gls{pc}. In fact, they do not have access to the other user's information. At the same time, both users can access $t$ common data streams through \gls{cc} channel.
In figure \ref{fig:sub1}, it can be observed an applicable scenario where users 1 and 2 are using a  broadcasting service simultaneously. They utilize \gls{cc} for shared information while each user also receives their private and personal information through \gls{pc}.

\subsection{Scenario \RNum{2}: $C\leq G $}
% Similar to scenario \RNum{1}, there are $s=C$ data streams and $\mathbf{V}$ is a square matrix. Also, the precoding matrix is defined as in scenario \RNum{1}. The input-output relation for the processed received signgabal is  represented by \eqref{eq:gsvd diagonal channel input output equation}. Matrices $\mathbf{\Sigma}_1$ and $\mathbf{\Sigma}_2$ are as follows

Similar to scenario \RNum{1}, there are \(s=C\) data streams, and \(\mathbf{V}\) is a square matrix. The precoding matrix is defined as in scenario \RNum{1}. The input-output relation for the processed received signal is given by \(\eqref{eq:gsvd diagonal channel input output equation}\). Matrices \(\mathbf{\Sigma}_1\) and \(\mathbf{\Sigma}_2\) are as follows

\begin{equation}
    \mathbf{\Sigma}_1 = 
    \begin{bmatrix}
    C_1\\
     \mathbf{0}_{r\times MNC}
    \end{bmatrix}, \mathbf{\Sigma}_2 = 
    \begin{bmatrix}
    C_2\\
     \mathbf{0}_{r\times MNC}
    \end{bmatrix},
    \label{eq:scenario 2 sigma}
\end{equation}
where $\mathbf{C}_1, \mathbf{C}_2\in\mathbb{C}^{MNC\times MNC}$ are  diagonal matrices with the \gls{cc} coefficients for user 1 and user 2.  In this scenario we have $r = G-C$. By comparing \eqref{eq:scenario 2 sigma} and \eqref{eq:gsvd diagonal channel input output equation} it can be noted  the first $C$ antennas in each receiver can be used to extract the information. Moreover, 
although the last 
$r$
 antennas in each receiver appear unused for data detection, they contribute to diversity gain. This gain is integrated into the precoding matrix $\mathbf{P}$, improving its performance in interference cancellation.
A broadcasting scenario where each user wants to have access to the same information, as illustrated in figure \ref{fig:sub2}.

% where the value of $\rho$ can be approximated by 
% \begin{equation}
%     \rho = \sqrt{\frac{1}{E\left\{\sum_{i=0}^{k}\sigma_i^{-1}\right\}}}
% \end{equation}
% where $\sigma_i$s are the eigenvalue of the matrix $\mathbf{H}\mathbf{H}^H$ where $\mathbf{H} = [\mathbf{H}_1^T,\mathbf{H}_1^T]^T$.

\subsection{Scenario \RNum{3}: $2G\leq C$}

In this case, the number of data streams is $s = 2G$, allowing us to allocate $s/2$ streams for each user. The matrix $\mathbf{V}$ is not square and can be decomposed as follows
\begin{equation}
     \mathbf{V}^H = [\mathbf{W}^H\mathbf{R}, \mathbf{0}_{MNs\times t}]\mathbf{Q}^H. \end{equation}
Here, $\mathbf{W} \in \mathbb{C}^{MNs \times MNs}$ and $\mathbf{Q} \in \mathbb{C}^{MNC \times MNC}$ are unitary matrices, $ t = C - 2G $, and $\mathbf{R} \in \mathbb{C}^{MNs \times MNs}$ is an invertible matrix with the singular values of the matrix $\mathbf{H}$.
The precoding matrix is defined as\footnote{For simulations, this simplifies to $\mathbf{P} = \mathbf{V} (\mathbf{V}^H \mathbf{V})^{-1}$, as GSVD beamforming simplifies to \gls{zf} transmission when the $2G\leq C$ \cite{Senaratne6415345}.}
$
     \mathbf{P} = \mathbf{Q}_{\left\{:,1:MNs\right\}} \mathbf{R}^{-1} \mathbf{W} \in \mathbb{C}^{MNC \times MNs} $.
 At the receiver, after applying the detection matrix for each user, we have 
 
\begin{equation}
    \resizebox{1\hsize}{!}{$\begin{cases}
\smash[t]{
 \mathbf{r}_1 = \mathbf{U}_1^H\mathbf{U}_1\mathbf{\Sigma}_1\overset{\mathbf{V}^H}{\overbrace{[\mathbf{W}^H\mathbf{R}, \mathbf{0}_{MNs\times t}]\mathbf{Q}^H}}\overset{P}{\overbrace{\mathbf{Q}_{\{:,1:MNs\}}\mathbf{R}^{-1}\mathbf{W}}}\mathbf{s}+ \mathbf{n}^{'}_1
} \\
\smash[b]{
\mathbf{r}_2 = \mathbf{U}_2^H\mathbf{U}_2\mathbf{\Sigma}_2[\mathbf{W}^H\mathbf{R}, \mathbf{0}_{MNs\times t}]\mathbf{Q}^H\mathbf{Q}_{\{:,1:MNs\}}\mathbf{R}^{-1}\mathbf{W}\mathbf{s}+ \mathbf{n}^{'}_2
}.
\end{cases}
$}
\label{eq:2G<C}
\end{equation}

The multiplication of  matrix $\mathbf{V}^H$ by  precoding matrix $\mathbf{P}$ is the identity matrix $\mathbf{I}_{s}$, so  \eqref{eq:2G<C} will reduce to \eqref{eq:gsvd diagonal channel input output equation}, with 
\begin{equation}
    \mathbf{\Sigma}_1 = 
    \begin{bmatrix}
    \mathbf{I}_{MNG}& \mathbf{0}_{MNG}     
    \end{bmatrix}, \mathbf{\Sigma}_2 = 
    \begin{bmatrix}
    \mathbf{0}_{MNG}&\mathbf{I}_{MNG}
    \end{bmatrix}.
    \label{eq:sigma 3rd scenario}
\end{equation}
Comparing \eqref{eq:sigma 3rd scenario} and \eqref{eq:gsvd diagonal channel input output equation}, it can be noted that each user can detect the information symbols solely through the \gls{pc}, with no common information transmitted via the \gls{cc}. Consequently, none of the users have access to each other's information, as illustrated in figure \ref{fig:sub3}.
% \section{Discussing Scenarios}
% In this section we are now considering different scenarios and analyzing the matrices $\mathbf{\Sigma}_1$, $\mathbf{\Sigma}_2$ and $\mathbf{V}$ and proposing an real world application for each of them.

% Consider $\sigma_{1,i}$ and $\sigma_{2,i}$ as the diagonal element of $\mathbf{\Sigma}_1$ and $\mathbf{\Sigma}_2$. So this values are realted via $\sigma_{1,i}^{2} + \sigma_{2,i}^{2} = 1$.
% All the cases are considered with $M = 16$ and $N = 8$ and consider we are in the reach scattering environment so the \gls{dd} channel matrix from antenna to antenna $\mathbf{H}_{g,i,c}$ is full rank.

% \subsection{G = 2, C = 4}
% In this case we have 
% $\mathbf{U}_1,\mathbf{U}_2\in\mathbb{C}^{2MN\times 2MN}$, $\mathbf{\Sigma}_1 =         [\mathbf{I}_{2MN},\mathbf{0}_{2MN}]$,  $ \mathbf{\Sigma}_2=[\mathbf{0}_{2MN},\mathbf{I}_{2MN}]$, and $\mathbf{V}\in\mathbb{C}^{4MN\times4MN}$.
% So each receiver only experience their own \gls{pc}.
% \subsection{G = 3, C = 4}
% In this case we have 
% $\mathbf{U}_1,\mathbf{U}_2\in\mathbb{C}^{2MN\times 2MN}$, $\mathbf{\Sigma}_1 =         [\mathbf{I}_{2MN},\mathbf{0}_{2MN}]$,  $ \mathbf{\Sigma}_2=[\mathbf{0}_{2MN},\mathbf{I}_{2MN}]$, and $\mathbf{V}\in\mathbb{C}^{4MN\times4MN}$.
% So each receiver only experience their own \gls{pc}.
% \subsection{G = 4, C = 4}
% \subsection{G = 5, C = 4}

\section{simulation results}

In this section, we evaluate the performance of the proposed \gls{gsvd} decomposition. The simulation parameters are listed in Table \ref{tab:simulation parameters}. The maximum speed of the environment is set to $v_{\text{max}} = 500$ km/h. Consequently, the relative maximum Doppler shift and the normalized Doppler shift are calculated as $\nu_{\text{max}} = \frac{v_{\text{max}} \times f_c}{\text{light speed}} = 1853$ Hz and $k_{\text{max}} = \nu_{\text{max}} T_f = 0.9883$, respectively. The information bits are mapped into information symbols using 4-\gls{qam} modulation. For each scenario, we compared our method with various \gls{mu}-\gls{mimo} schemes. In scenario \RNum{1}, we used the \gls{bd} precoding scheme proposed in  \cite{Bandemer4022531} followed by \gls{mmse} equalizer at the receiver. In scenario \RNum{2}, we employed an \gls{mmse} equalizer at the receiver. Lastly, in scenario \RNum{3}, we considered \gls{mmse} precoding.

% Fig.~\ref{s1} represents the \gls{ber} curve versus \gls{snr}. The results are shown for Scenario \RNum{1}, where 
% $G\leq C\leq 2G$, with $C = 4$ and $G=3$. As mentioned in Section \RNum{3}, this scenario includes both \gls{pc} and \gls{cc} for the \gls{gsvd}-based \gls{otfs} system. The simulation results indicate a significant gap between \gls{pc} and \gls{cc}, with \gls{pc} demonstrating superior performance. Comparing our proposed method with the \gls{bd} precoding followed by an \gls{mmse} equalizer, we observe that while the \gls{bd}-\gls{mmse}  approach exhibits better performance, it comes at the cost of being able to transmit only two data streams (one for each user), thus limiting the transmission of common information.

\begin{figure}[t]
\centerline{\includegraphics[width=0.49\textwidth]{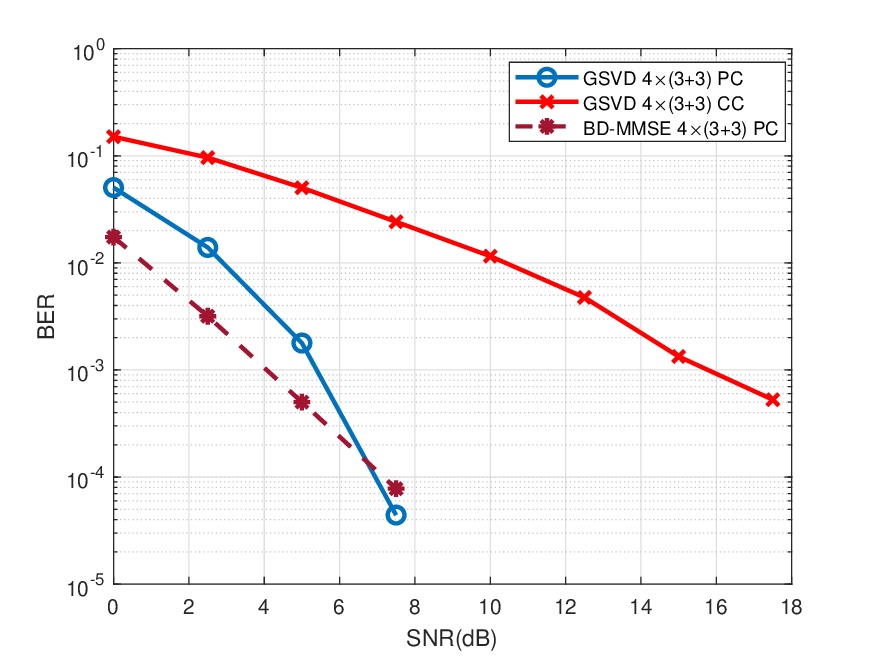}}
\caption{\gls{ber} vs \gls{snr} for Scenario \RNum{1}, comparing \gls{gsvd}-based precoding and \gls{mmse} precoding.}
\label{s1}
\end{figure}

Fig.~\ref{s1} represents the \gls{ber} curve versus \gls{snr}. The results are shown for Scenario \RNum{1}, where \( G \leq C \leq 2G \), with \( C = 4 \) and \( G = 3 \). As mentioned in Section \RNum{3}, this scenario includes both \gls{pc} and \gls{cc} for the \gls{gsvd}-based \gls{otfs} system. The simulation results indicate a significant gap between \gls{pc} and \gls{cc}, with \gls{pc} demonstrating superior performance. In the \gls{gsvd} system, there is 1 \gls{pc} and 2 \glspl{cc}, whereas in the \gls{bd}-\gls{mmse} approach, there are 2 \glspl{pc}. Comparing our method with \gls{bd} precoding and \gls{mmse} equalizer, we observe better performance with the \gls{bd}-\gls{mmse} approach, but it only transmits two data streams, limiting common information transmission.

\begin{figure}[t]
\centerline{\includegraphics[width=0.49\textwidth]{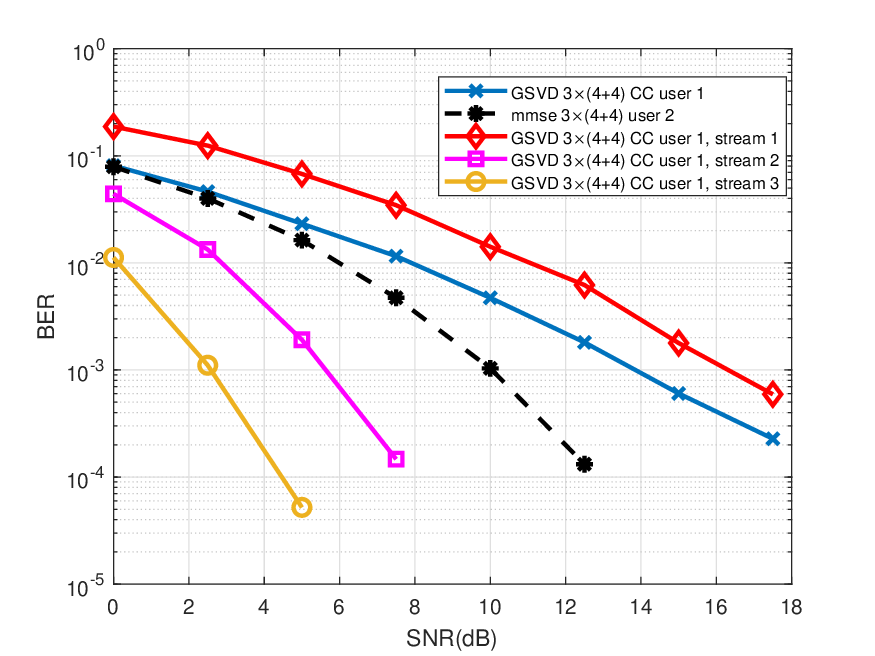}}
\caption{\gls{ber} vs \gls{snr} for Scenario \RNum{2}, comparing \gls{gsvd}-based precoding and \gls{mmse} equalization.}
\label{s2}
\end{figure}

 Fig.~\ref{s2} illustrates Scenario \RNum{2}, where 
\(C \leq G\)
and \(C = 3\), \(G = 4\). In this scenario, only \gls{cc} is present. The \gls{ber} performance of the \gls{gsvd}-based \gls{otfs} system for each user is inferior to \gls{mmse} equalization. This is attributed to the fact that some \gls{cc} coefficients are quite small. However, when analyzing each data stream individually rather than collectively, it is observed that the \gls{ber} for two of the streams surpasses that of \gls{mmse} equalization, while one stream performs worse than \gls{mmse}. This suggests that the proposed scheme can perform effectively with the application of an appropriate coding scheme such as low-density parity check (LDPC). Furthermore, the \gls{gsvd}-based method offers the advantage of channel equalization at the receiver using a single-tap equalizer, thereby transforming the channel matrix into a diagonal form. This simplifies the process significantly, as handling large non-diagonal matrices in \gls{otfs} modulation has always been challenging.

\begin{figure}[t]
\centerline{\includegraphics[width=0.49\textwidth]{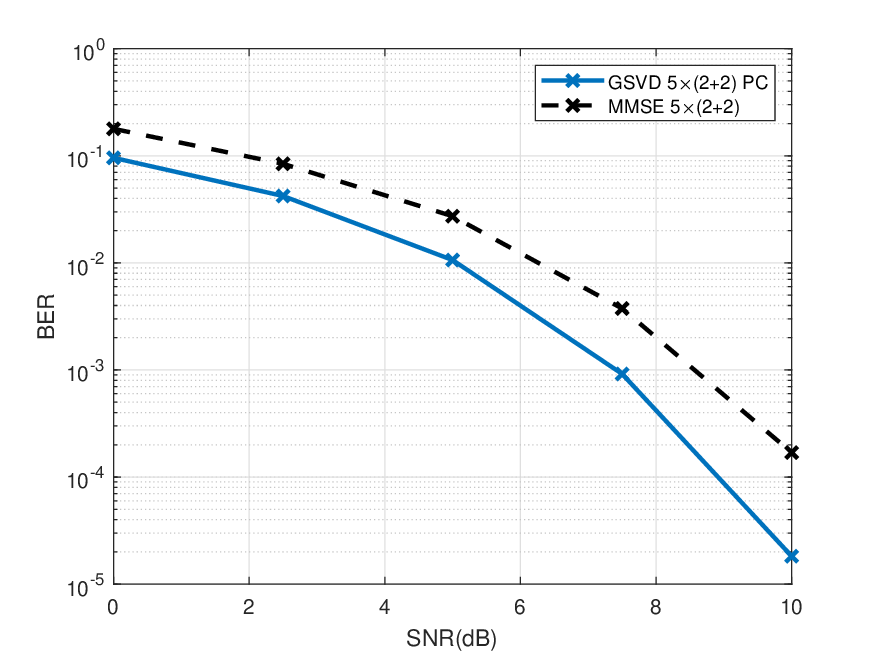}}
\caption{\gls{ber} vs \gls{snr} for Scenario \RNum{3}, comparing \gls{gsvd}-based precoding and \gls{mmse} precoding.}
\label{s3}
\end{figure}

 Fig.~\ref{s3} illustrates Scenario III, where \( 2G \leq C \) and \( C = 5 \), \( G = 2 \). In this scenario, our proposed method outperforms the \gls{mmse} precoding technique. It is evident that in this scenario, we only have \gls{pc}. As with other scenarios, the performance of User 1 and User 2 for both \gls{mmse}-based precoding and \gls{gsvd}-based precoding remains the same.

\begin{figure}[t]
\centerline{\includegraphics[width=0.49\textwidth]{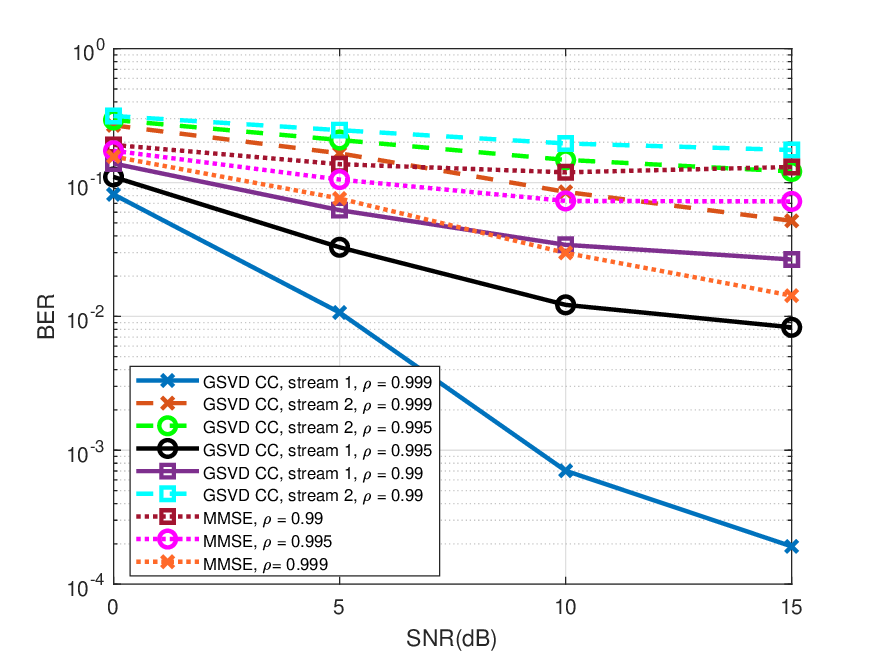}}
\caption{\gls{ber} vs \gls{snr} for Scenario \RNum{2}, comparing \gls{gsvd}-based precoding and \gls{mmse} equalization in the presence of estimation error.
}
\label{esterror}
\end{figure}

In Fig.~\ref{esterror}, we show the system's performance with channel estimation errors. With \( C = 2 \) and \( G = 2 \), this case fits either Scenario \RNum{1} or \RNum{2}.
If the time domain channel coefficient is \( h_k \), we model the estimated channel coefficient using \( h_{k,\text{est}} = \rho h_k + \epsilon \), where \( 0 \leq \rho \leq 1 \) and \( E\{|\epsilon|^2\} = (1 - \rho^2)E\{|h_k|^2\} \) (with \( E\{\cdot\} \) indicating the expected value). The figure shows the superiority of the proposed method compared to \gls{mmse} equalization in the presence of channel estimation errors for the stream with better channel conditions, and performance almost similar to the \gls{mmse} method for the stream with worse conditions.

\begin{table}[t]
\caption{Simulation parameters}
\begin{center}
\begin{tabular}{|c|c|c|}
\hline
\textbf{ Parameters}&\textbf{Values} \\ \hline
Carrier frequency & $f_c=4$ GHz \\
\hline
Sub-carrier spacing& $\Delta_f=15$ kHz  \\
\hline
$(M,N)$ & $(16,8)$\\ \hline
Tap delays (ns)& $[0, 30, 150,310,370,710,1090,1730,2510]$\\ \hline
Tap  powers (dB)& $-[0, 1.5, 1.4,3.6,0.6,9.1,7,12,16.9]$\\ \hline
Max. speed &  $500$ km/h \\ \hline

\end{tabular}
\label{tab:simulation parameters}
\end{center}
\end{table}

\section{conclusion and research opportunities}
In this paper, we proposed the adoption of the \gls{gsvd} decomposition for a downlink two-user \gls{mimo}-\gls{otfs} system and derived analytical expressions for three scenarios. The system is based on \gls{gsvd} decomposition of each user's channel and on applying a precoding and detection matrix.  The simulation results indicate the superiority of the proposed \gls{gsvd} scheme compared to conventional \gls{mmse}.
The extension of the \gls{gsvd} for two-user into \gls{hogsvd} for more users in uplink and downlink scenarios can be an interesting future work path.

\bibliography{conference_101719}
\bibliographystyle{IEEEtran}
% \begin{thebibliography}{00}
% \bibitem{b1} G. Eason, B. Noble, and I. N. Sneddon, ``On certain integrals of Lipschitz-Hankel type involving products of Bessel functions,'' Phil. Trans. Roy. Soc. London, vol. A247, pp. 529--551, April 1955.
% \bibitem{b2} J. Clerk Maxwell, A Treatise on Electricity and Magnetism, 3rd ed., vol. 2. Oxford: Clarendon, 1892, pp.68--73.
% \bibitem{b3} I. S. Jacobs and C. P. Bean, ``Fine particles, thin films and exchange anisotropy,'' in Magnetism, vol. III, G. T. Rado and H. Suhl, Eds. New York: Academic, 1963, pp. 271--350.
% \bibitem{b4} K. Elissa, ``Title of paper if known,'' unpublished.
% \bibitem{b5} R. Nicole, ``Title of paper with only first word capitalized,'' J. Name Stand. Abbrev., in press.
% \bibitem{b6} Y. Yorozu, M. Hirano, K. Oka, and Y. Tagawa, ``Electron spectroscopy studies on magneto-optical media and plastic substrate interface,'' IEEE Transl. J. Magn. Japan, vol. 2, pp. 740--741, August 1987 [Digests 9th Annual Conf. Magnetics Japan, p. 301, 1982].
% \bibitem{b7} M. Young, The Technical Writer's Handbook. Mill Valley, CA: University Science, 1989.
% \end{thebibliography}

\end{document}